# Artificial intelligence in deep brain stimulation for movement disorders: a systematic review and technology readiness assessment

Zohra Souei[#1,2], Muhammad Mushhood Ur Rehman[#3], Harith Akram[4], Jocelyne Bloch[5,6], Stephan Chabardes[7], Alfonso Fasano [8, 9], Marwan Hariz[10, 11], Joachim K. Krauss[12], Andrea A. Kühn[13], Patricia Limousin[4, 14], Daniel E. Lumsden[15, 16], Eduardo M. Moraud [6,17], Elena Moro[18], Martin M. Reich[19], Ludvic Zrinzo[4], Xavier Vasques[†2], Laura Cif [†2,20]

## Affiliations:

[1] Department of Neurosurgery, Military University Hospital of Sfax, Tunisia

[2] Institut du Neurone, Montferrier sur Lez, France

[3] Edinburgh Medical School, University of Edinburgh, Scotland, UK

[4] Unit of Functional Neurosurgery, UCL Queen Square Institute of Neurology, London, UK

[5] Department of Neurosurgery, Lausanne University Hospital (CHUV), Switzerland

[6] NeuroRestore, Ecole Polytechnique Fédérale de Lausanne, University Hospital Lausanne and University of Lausanne, Switzerland

[7] Department of Neurosurgery, CHU of Grenoble, Grenoble, France

[8] Department of Biomedical Sciences, Humanitas University, Via Rita Levi Montalcini 4, 20090 Pieve Emanuele, Milan, Italy

[9] IRCCS Humanitas Research Hospital, via Manzoni 56, 20089 Rozzano, Milan, Italy

[10] Department of Clinical Science, Neuroscience, Umeå University, 90185 Umeå, Sweden

[11] UCL Institute of Neurology, Queen Square, Box 146, WC1N 3BG London, UK

[12] Department of Neurosurgery, Hannover Medical School, Germany

[13] Department of Neurology, Charité- Universitätsmedizin Berlin, Berlin, Germany

[14] Department of Clinical and Movement Neurosciences, University College London, London, UK.

[15] Complex Motor Disorder Service, Evelina London Children's Hospital, London, UK

[16] Research Department of Early Life Imaging, Biomedical Engineering and Imaging Sciences, King's College London, UK

[17] Neuro-X Institute, Ecole Polytechnique Fédérale de Lausanne, Geneva, Switzerland

[18] Grenoble Alpes University, CHU of Grenoble, Division of Neurology, Grenoble Institut of Neurosciences, Grenoble, France

[19] Department of Neurology, Julius Maximilians University Hospital Würzburg, Germany

[20] Neurology Unit, Department of Clinical Neurosciences, Lausanne University Hospital (CHUV) and University of Lausanne (UNIL), Lausanne, Switzerland

# These authors contributed equally to this work
† These authors contributed equally to this work

## Corresponding author:

Laura Cif: cifanalaura@gmail.com

# Abstract

Artificial intelligence (AI) is increasingly explored across deep brain stimulation (DBS) for movement disorders, yet whether current systems are approaching deployment remains unclear. To characterise their scope, validation maturity, and translational readiness, we systematically evaluated 239 peer-reviewed studies published between 2000 and 2025, assessing AI methods, validation practices, and barriers constraining clinical translation.

Research was dominated by Parkinson's disease and subthalamic nucleus targeting, with limited coverage of other disorders and targets. Most studies reported encouraging internal performance; however, external validation was rare, evaluations remained predominantly retrospective and single-centre, and more than one-quarter involved small-sample, high-dimensional datasets with elevated overfitting risk. Technology readiness assessment revealed that most systems remain at early-to-intermediate translational stages, constrained more by limited validation than by algorithmic inadequacy, compounded by the biological heterogeneity and dynamic complexity inherent to DBS. Nevertheless, emerging external and prospective studies suggest a field moving toward clinical maturity, with promising applications in targeting, programming, outcome prediction, and adaptive therapy delivery.

# Introduction

Deep brain stimulation (DBS) is an established therapy for Parkinson's disease (PD) and may also be considered for carefully selected patients with dystonia or essential tremor (ET) [1–4]. However, DBS remains complex and resource-intensive, requiring nuanced patient selection, precise anatomical targeting, and iterative postoperative programming. Clinical outcomes vary substantially across individuals, reflecting heterogeneity in disease phenotype, neuroanatomy, and stimulation response that is not readily captured by conventional decision-making frameworks. Artificial intelligence (AI) has been proposed to support decision-making across the DBS pathway [5,6]. Using multimodal data, these methods may improve prediction beyond manual integration alone [7].

Applications include patient stratification and outcome prediction, automated analysis of neural signals to infer circuit states, and data-driven programming strategies to personalize stimulation [8,9]. Computational modelling may further link clinical phenomenology to circuit-level mechanisms, helping identify new targets and stimulation strategies, particularly in dystonia, where pathophysiology remains incompletely understood [10]. These capabilities are especially relevant as DBS evolves toward feedback-informed and adaptive systems operating across fluctuating clinical states [11,12]. Despite this promise, clinical translation remains difficult, requiring not only robust validation but also compliance with evolving regulatory frameworks for high-risk medical AI, including requirements for transparency, human oversight, and post-market monitoring [13,14].

Prior systematic reviews of AI in DBS for movement disorders (MDs) have been limited in scope or translational assessment. Earlier analyses identified persistent challenges, including small cohorts and inadequate validation across 73 machine learning (ML) studies [15]. Others focused on specific subdomains, such as adaptive stimulation [16], or broader DBS applications [17], but did not assess validation maturity, translational readiness, or barriers to clinical integration across the full workflow, nor frame findings in terms of deployment readiness rather than algorithmic performance alone. As a result, it remains unclear whether these methodological limitations have been addressed and whether current systems are approaching deployment readiness rather than remaining laboratory proof-of-concept tools. Many models still rely on small, retrospective, single centre datasets with limited external validation, increasing susceptibility to overfitting and inflated performance estimates, particularly when patient-level separation and leakage control are not rigorously enforced [18]. Moreover, high reported accuracy is often presented without calibration, uncertainty quantification, or clinically meaningful metrics needed for high-stakes deployment, where false positives and negatives can have unequal consequences. Clinicians managing DBS patients daily have access to few AI tools that meaningfully alter what they can offer, yet the corpus is expanding faster than the field's capacity to appraise it critically, making a systematic assessment of what constrains generalisability and deployment readiness both timely and necessary.

The primary question this review addresses is whether AI in DBS for MDs is converging toward clinical deployment or accumulating technically refined work that remains detached from the clinical problem it claims to solve. This perspective distinguishes the present review from prior work in three respects: in scope, by spanning the full DBS clinical trajectory rather than a single workflow stage or condition; in analytical framework, by mapping validation practices and governance dimensions rather than cataloguing methods and performance alone; and in translational grounding, by applying adapted Technology Readiness Level scoring across all 239 included studies to characterize where the corpus actually sits on the laboratory-to-clinic continuum.

To address these gaps, this systematic review addresses four questions: (1) which AI methods and data modalities have been applied across DBS workflow stages, including patient selection, targeting, programming, adaptive stimulation, and outcome assessment; (2) whether current validation practices, including external, temporal, and nested cross-validation support claims of clinical generalizability; (3) the translational maturity of existing AI-DBS systems; and (4) which methodological, regulatory, and infrastructure barriers constrain clinical deployment.

# Results

## Study characteristics

We included 239 peer-reviewed studies published between 2000 and 2025 (median publication year, 2021) applying AI to DBS for MDs (***Figure 1***); 70.3% (n= 168) were published from 2020 onwards. Studies originated from 30 countries, led by the United States (n= 64, 26.8%), followed by China (n= 52, 21.8%) and Germany (n= 27, 11.3%). Among studies reporting funding **(**n= 192**),** support was predominantly derived from government agencies (154/192, 80.2%)**,** frequently in combination with other sources. Academic or institutional funding was reported in 56 studies (29.2%) and non-profit organizations (49/192, 25.5%). Industry support was less common (24/192, 12.5%) and rarely occurred alone (1/192, 0.5%), appearing mainly within mixed funding arrangements.

PD predominated (79.91%), followed by mixed cohorts (8.78%), ET (7.11%), and dystonia (4.18%). The subthalamic nucleus (STN) was the most frequently studied DBS target (61.5%), followed by mixed or multiple targets (18.4%). The globus pallidus internus (GPi) and the ventral intermediate nucleus of the thalamus (Vim) were each reported in 5.85% of studies.

Human cohorts accounted for 84.1% of studies (n= 201/239; median n= 20, IQR 10–50, range 1–770), whereas 10.0% used computational models and 4.2% animal models. Two studies (0.8%) used mixed designs and two (0.8%) used semi-synthetic data.

The main study-level characteristics of the included articles are summarized in ***Table 1***. A complete list of included studies and extracted data is available in a public repository on Figshare (doi: 10.6084/m9.figshare.31750372)

## AI model architecture across the DBS workflow

AI applications were most frequently directed toward stimulation optimisation, with adaptive DBS (aDBS) (n= 73) and programming (n= 27) together accounting for 41.8% of all studies. Other applications included targeting (n= 62) screening and outcome prediction (n= 36), mechanistic (n= 19), and objective and automated assessment (n= 19); a small number addressed other uses (n= 3).

Model architecture usage is summarized in ***Table 2*** and ***Figure 2***. Many studies used multiple models to compare architectures for a single task or support different pipeline steps.

Supervised learning predominated (n= 210 studies), while model selection varied across workflow stages. Outcome prediction primarily relied on classical machine-learning approaches, including linear and generalized linear models (Linear/GLM, n= 19), kernel-based models such as support vector machines (Kernel/SVM, n= 22), and tree-based or ensemble models (Tree/Ensemble, n= 18).

Targeting employed both classical approaches—kernel-based models (n= 16) and Linear/GLM models (n= 12)—and deep learning (DL) methods, particularly convolutional neural networks (CNNs, n= 24), reflecting the prominence of image-guided localization tasks.

Adaptive DBS (aDBS) studies showed substantial architectural diversity, including Linear/GLM models (n= 24), kernel-based models (n= 23), tree-based or ensemble models (n= 17), and feedforward or representation-learning models (n= 18). Programming studies most commonly used Linear/GLM (n= 10) and kernel-based models (n= 9). Mechanistic studies most frequently employed linear and generalized linear models (n= 7) and probabilistic or Bayesian approaches (n= 6)**,** with other architectures used less commonly.

Explainability methods (XAI) were reported in 66.1% of studies, with substantial variation across workflow stages. Feature importance (n= 94) and intrinsically interpretable models (n= 72) represented the most commonly reported approaches. Post-hoc approaches remained rare (5.43%). Some studies used multiple XAI methods (***Figure 3 A, B***). XAI use was highest in outcome prediction, which was also the only stage with notable SHAP implementation (n=7/8 total SHAP studies). Mechanistic studies ranked second and more often favoured intrinsic interpretability. Clinician-oriented decision-support interfaces were described in 12.9% including nomograms, intraoperative signal-visualization and real-time classification tools, anatomical planning overlays, imaging-based risk heat maps, and web-based calculators.

## Validation landscape

Validation relied predominantly on internal resampling. Cross-validation was the most common approach, including k-fold cross-validation (30.9%), leave-one-out cross-validation (LOOCV, 9.62%), and leave-one-subject-out cross-validation (LOSO, 8.79%). Hold-out validation (single train-test split) was used in 20.92% of studies. Nested cross-validation, which separates hyperparameter tuning from performance evaluation to reduce optimistic bias, was reported in 7.11%. Simulation-based validation appeared in 7.95%, mainly in aDBS and programming studies. Validation strategies varied across workflow stages (**Table 3*****, Figure 3C***). Across workflows, 26.4% of studies (n= 63) operated in small-sample, high-dimensional settings in which the number of features exceeded the number of training samples, increasing the risk of overfitting in the absence of adequate regularization or validation. Accuracy was most common (56.1 %), followed by sensitivity (31.4 %), AUC (28.9%), and specificity (25.9%). More informed metrics were less frequent, including confusion matrices (13.8%). F1-score (13.4%) and precision (12.13%).
Metric choice varied by workflow: classification metrics dominated objective assessment (94.7%) and targeting (82.3%), whereas regression metrics were more common in outcome prediction (50.0%) followed by objective assessment (36.8%) and programming (33.3%). Dedicated segmentation metrics (for example Dice, Hausdorff, IoU) appeared in only 12.9% of targeting-related metric mentions despite image-based tasks. Most targeting studies instead relied on classification-oriented metrics, often reporting accuracy at the voxel or region level rather than evaluating spatial overlap or boundary precision.

## TRL distribution and translational gaps

TRL analysis revealed a strong concentration at early development stages (***Figure 4A***): 86.2% of studies were classified as TRL 2–4, 13.8% as TRL 5–6, and none reached routine clinical deployment (TRL ≥7). The most common level was TRL 4 (65.7%). Attainment of TRL 5–6 varied across workflow stages, reaching 21% (n=4/19) in objective and automated assessment, 19.2% (n=14/73) in aDBS, 18.5% (n=5/27) in programming, 11.3% (n=7/62) in targeting, 5.6% (n=2/36) in screening and outcome prediction, and 5.3% (n=1/19) in mechanistic studies.

Within the aDBS category, TRL 6 was confined to tremor in ET across motor control applications and to sleep modulation in a single PD study; motor state control in PD, the largest body of aDBS work, did not exceed TRL 5 in any signal modality. (***Figure 4B***).

Translation readiness indicators are summarized in **Table 4**. Regulatory frameworks were referenced in 5.02% of studies, while model lifecycle management considerations appeared in only 2.5%. Multi-centre data or collaboration was reported in 14.64%, and open-source code, models, or data in 20.9%. Prospective evaluation was reported in 6.27% of studies. Clinical workflow integration or decision-support interfaces were mentioned in 51.04%.

Workflow stages differ systematically in translational readiness across the dimensions most relevant to clinical generalizability. Validation strategies, designed to assess generalisability beyond the development dataset, were uncommon across the corpus (***Table 3, Figure 3D***). External validation, defined as testing on independent cohorts from different institutions, is reported in 5.86% of studies overall, peaking in programming at 14.8% (n=4/27) and lowest in adaptive DBS at 2.7% (n=2/73) despite the largest study volume in that stage. Temporal and prospective evaluation remain rare (2.5% and 5.4% overall respectively), concentrated in programming and adaptive DBS. Robustness to site or data-source shift, formally assessed in the externally validated studies summarised in ***Supplementary Table 1***, follows a task-dependent pattern: categorical tasks with anatomical ground truth (contact selection, STN localization) preserve cross-institutional accuracy, whereas tasks generalizing complex biological or behavioural signals (outcome prediction, cross-vendor MER classification) degrade substantially across cohorts. Practical deployment constraints are reflected in TRL attainment, with no study reaching TRL ≥7; TRL 5–6 concentrates in objective assessment (4/19) and adaptive DBS (14/73), while workflow integration narratives are highest in targeting (43/62) and outcome prediction (18/36) but rarest in adaptive DBS (29/73). These cross-stage differences are summarized in ***Table 3***.

# Discussion

AI now spans the full DBS workflow, with broad momentum across methods and data modalities. Yet breadth of application has not translated into robust validation: performance reported under controlled conditions is still rarely demonstrated across centres, devices, and patient populations. Where AI is most likely to deliver near-term clinical value can be located along three dimensions across the workflow stages mapped in ***Table 3***: the acuity of the underlying clinical need, the maturity of technical validation achieved, and the clarity of the clinical integration pathway. Targeting and programming converge favourably on all three; outcome prediction, adaptive DBS, objective assessment, and mechanistic modelling each occupy distinct intermediate positions on this landscape, as the paragraphs that follow develop in detail. Bridging this gap will require not only better algorithms but also shared understanding of what different AI approaches can realistically deliver in DBS, together with more rigorous evidence generation and reporting.

One area in which AI may offer genuine clinical value is screening and outcome prediction. By integrating heterogeneous clinical, imaging, neurophysiological, and connectivity data, AI can formalize forms of implicit clinical synthesis that are difficult to reproduce consistently, particularly in dystonia, where both aetiology and phenotypic heterogeneity complicate surgical decisions. However, evidence remains limited: among 36 outcome-prediction studies, only two addressed dystonia. In one paediatric cohort (n=96), decision trees identified central motor conduction time and baseline severity as predictors of outcome (66-67% accuracy), whereas in adults (n=32), SVM predicted response (11±7% prediction error) using early motor improvement, phenomenology characteristics (phasic 67% vs tonic 33%), body distribution, and VTA-derived measures [19 20]. In PD, multimodal approaches have shown mixed added value, with some studies identifying single markers, such as nucleus basalis of Meynert volume ($R^2$=0.149, n=55) or higher beta activity, as stronger predictors than broader clinical variables [21] [22]. These tools may help convert subjective clinical judgment into more explicit and testable risk estimates, supporting counselling and resource allocation [23].

In targeting, this convergence is anchored in the directness of the clinical need (***Table 3***): standardization, speed, and reduced operator dependence are immediately relevant to surgical practice. Manual MER analysis suffers inter-observer variability and experience-dependence; AI can provide more reproducible signal classification and near-real-time feedback, including short signal segments, with potential to reduce human errors, reduce procedural burden, reduce infection risk and improve patient comfort [24–27]. DL models detect non-linear patterns (high-frequency oscillations, beta-band dynamics) difficult for humans to distinguish visually or audibly [25,28], identifying STN "sweet spots" correlating with optimal motor outcomes rather than just anatomical borders [29,30]. HMM and Bayesian approaches utilize trajectory history to improve accuracy through spatial continuity constraints [31–33]. However, while traditional ML studies systematically compared feature sets, the recent shift toward end-to-end DL tends to bypass explicit feature engineering in favour of "black-box" extraction from raw data or spectrograms [24,30,34]. Beyond electrophysiology, AI enables direct patient-specific anatomical/structure segmentation in native space, capturing individual variability while avoiding atlas registration errors. Models like GP-net demonstrates automated subject-specific delineation outperforms atlas-based targeting [35]. Neural networks predicting pyramidal tract side effects (PTSE) from electrode coordinates, current values, and trajectory angles in GPm-DBS achieved kappa=0.78 (n=10, 988 tests), enabling preoperative PTSE risk visualization and trajectory selection maximizing therapeutic windows-capabilities difficult to formalize through expert judgment alone relying on iterative intraoperative testing [36]. The pedunculopontine nucleus, a brainstem target explored for refractory gait instability in PD, exemplifies this evolution. Patient-specific MRI localization addressed the targeting variability of atlas-based approaches highlighting the contribution of direct MRI-based localization of the pedunculopontine nucleus, though limited resolution and contrast at 1.5T still required inference from surrounding white matter tracts [37]. Self-supervised DL for MRI super-resolution combined with quantitative susceptibility mapping demonstrates feasibility for direct delineation of the nucleus itself on individual patient scans [38]. AI-driven analysis may identify novel targets based on disease-specific structural changes : computational thalamic atrophy analysis in blepharospasm nominated ventral lateral anterior thalamus as candidate target [39]. More broadly, Image Quality

Transfer (IQT) frameworks, which enhance MRI resolution and contrast by learning mappings between acquisition protocols, are being integrated into commercial MR sequences, offering general-purpose image enhancement applicable to DBS targeting [40]. AI thus augments, not replaces, surgical expertise by standardizing targeting decisions and optimizing accuracy.

AI-assisted programming may improve access by reducing reliance on highly specialized expertise. In dystonia, ML-guided programming achieved greater symptom reduction than expert clinical programming (74.9% versus 63.3%; $p=0.012$), while SVM classifiers maintained high accuracy across independent institutions [41,42]. These tools enable high-quality programming outside specialized centres, reducing patient travel burden and financial barriers while personalizing therapy to individual response patterns [43]. Compressing optimization timelines shortens the typical one-year titration period: Bayesian optimization identifies optimal settings in 15-18 iterations versus 25 for exhaustive grid searches, while fMRI-guided ML achieved 76% accuracy predicting optimal parameters in treatment-naïve patients, potentially reducing programming to single sessions [41,44–46]. Directional leads introduce $10^{10}$-$10^{30}$ possible configurations, creating a combinatorial explosion that makes manual exploration impractical. AI navigates this complexity, making advanced lead designs feasible for routine clinical use [47,48]. Coordinating stimulation across multiple targets may further support personalized therapy. In simulated PD models (n=5), RL controllers managing simultaneous STN-GPi stimulation improved tremor suppression by ~80% while reducing energy consumption relative to single-target approaches, suggesting a computational framework for more efficient exploration of parameter combinations [49]. Multi-objective optimization algorithms help balance competing therapeutic goals by identifying parameter sets that define trade-offs between symptom control and side-effect risk. This eliminates 61% of the search space by discarding settings that are strictly inferior (worse on all dimensions), focusing programming exclusively on genuine therapeutic trade-offs [50]. Safe Bayesian optimization avoids side-effect-inducing parameters, reducing adverse events from 12% to 5% in dystonia despite training solely on efficacy data [44]. These systems function as decision-support tools navigating complexity while requiring continuous clinician oversight.

Among AI applications in DBS, adaptive stimulation represents the most technologically mature domain (***Figure 5 A***). Real-time classifiers, predominantly using STN LFP features across multiple small human cohorts, reliably detect motor states and stimulation-relevant biomarkers, addressing fundamental limitations of continuous stimulation through three mechanisms that deliver measurable patient-centered benefits. AI-enabled anticipatory control may allow stimulation to shift from reactive to pre-emptive adjustment: in movement-based aDBS, neural classifiers have predicted movement within 2 seconds in advance, enabling stimulation before tremor onset.[51–53]. Activity-aware approaches can also modulate stimulation according to behavioural context, potentially reducing the side-effect impact associated with fixed-parameter programming. Adaptive systems deliver higher stimulation during motor-demanding tasks while reducing amplitude during speech or cognitive activities. Emerging work demonstrates real-time decoding of gait events, freezing episodes, and force modulation from subthalamic LFPs in freely walking patients using commercial implants [54], addressing critical unmet needs as gait dysfunction remains poorly responsive to conventional DBS. Recent movement-responsive aDBS using remotely optimized decoders improved typing speed and reduced dyskinesia versus conventional DBS, maintaining efficacy over 14 months, validating behaviour-contingent stimulation and ML-based parameter optimization in home settings [55]. These context-dependent adjustments translate directly to reduced adverse events: by stimulating only during symptomatic periods and modulating amplitude proportionally to symptom severity, AI-based systems minimize current spread to adjacent structures, potentially reducing stimulation-related side effects including speech impairment, gait disturbances, and cognitive effects [56–67]. Although evidence for clinical superiority of aDBS over conventional continuous stimulation remains limited, particularly in the absence of large, long-term comparative trials, work in adaptive and automated DBS has clarified key principles for biomarker-driven optimization, parameter search, and scalable programming. As highlighted in recent reviews, these advances primarily address inefficiencies and variability in current DBS workflows and are likely to be essential for the clinical translation of increasingly complex stimulation technologies, with implications that extend beyond near-term improvements in symptom control [68].

AI-based objective and automated assessment addresses structural gaps in DBS care. Conventional scales provide episodic subjective snapshots incompletely capturing symptom variability and subtle changes; ML and DL methods extract motor signatures difficult to grade reproducibly or quantify real-time. Multimodal sensing (IMU, force sensors, mechanomyography) detects severity changes when UPDRS scores remain unchanged [69]. Digitized drawing tasks identify stimulation-dependent motor changes within seconds during programming, providing high-dimensional characterization [70]. Higher-order spectral methods reveal nonlinear tremor features (coupling strength, system complexity) not captured by power spectral analyses [71]. A particularly distinctive contribution is capturing dimensions invisible to standard scales and predicting therapeutic response: vision-based frameworks show baseline kinetic tremor amplitude independently predicts DBS outcome, not reflected in conventional ratings [72]. Video-based DL in cervical dystonia identifies kinematic and oscillatory features showing larger DBS effect sizes than static measures, functioning as outcome measures and prognostic indicators [73]. Practical utility is enhanced by the ability of these methods to operate on standard, low-cost devices rather than specialized laboratory infrastructure. Tablet-based tracking and smartphone-based support vector machine models can estimate tremor severity and detect DBS ON-OFF states with high temporal resolution using only embedded sensors [74]. Automated video-based hand-tracking achieves precision comparable to three-dimensional motion-capture systems (within ~2·6 mm for amplitude), indicating that routine clinical videos are sufficient for accurate kinematic analysis [72]. A particularly compelling use case arises in complex, mixed phenotypes in which multiple symptoms coexist, scenarios that demand expert assessment often unavailable across centres but are amenable to standardized, AI-assisted phenotyping [75]. These tools serve a dual role: beyond their direct clinical utility, they reduce label noise, strengthen training ground truth, and provide the deployment-relevant endpoints against which all downstream AI models must be validated. Nonetheless, clinical deployment requires validation in free-living conditions, multicentre evaluation under real-world constraints, assessment in mixed phenotypes, and seamless integration into clinical workflows. At present, the principal value of AI lies in mechanistic insight and task-specific monitoring in controlled settings rather than in generalizable prediction of individual outcomes across centres. For routine deployment, future frameworks will also need to incorporate privacy-preserving approaches, such as AIoT-based or federated learning architectures, to ensure secure handling of electronic health records during cloud-enabled processing.

AI-based mechanistic approaches have identified symptom-related neural signatures and circuit relationships that extend beyond classical rate-based models of basal ganglia dysfunction [76]. These findings suggest that MDs may be better understood in terms of dynamic patterns of network synchrony and state transitions rather than firing rates alone [77,78]. In particular, machine-learning approaches, including hidden Markov models, have linked different motor features to distinct large-scale oscillatory states, including persistent beta synchrony in akinesia and rigidity, shifts toward more prokinetic rhythms after dopaminergic therapy or STN-DBS, and changes in network-state stability associated with tremor and gait improvement [77,78]. In dystonia, ML analyses of pallidal recordings reveal asymmetric firing rates and burst dynamics challenging uniform abnormal firing-rate models. RF classifiers show GPi firing rates significantly elevated ipsilateral to head rotation direction in cervical dystonia, supporting network-based interpretation where symptoms arise from asymmetric feedback across interconnected circuits (basal ganglia, cerebellar pathways) rather than focal lesions [76]. Physiological markers extend beyond pallidal theta power to subthalamic-pallidal beta power coupling and theta phase coupling, both correlating significantly with clinical severity [79]. Optogenetic activation of cuneiform nucleus glutamatergic neurons controlled locomotor speed and reduced freezing in parkinsonian mice [80], illustrating how animal models combined with DL-based movement quantification reveal alternative DBS targets. ML-based single-cell transcriptomics and connectomics could systematically accelerate such discoveries through unbiased network mapping [81]. However, mechanistic insights remain exploratory with substantial translation barriers. Most findings derive from small single centre cohorts (typically 5-27 participants) limiting generalizability; model-based predictions often lack direct experimental validation and rely on normative rather than patient-specific connectomes. Current ML frameworks should be regarded as hypothesis-generating tools rather than definitive disease pathophysiology evidence.

Study design characteristics are consistent across workflows: predominantly single centre cohorts, small sample sizes, and near-complete absence of public benchmarks preventing reproducible comparison. Dystonia (n=10 exclusive studies, n=10 in mixed cohorts) remains underrepresented despite clinical burden and unique translational opportunities. Unlike PD/ET, dystonia exhibits more delayed therapeutic effects (weeks-to-months) and heterogeneous phenotype-specific outcomes, demanding AI approaches accommodating longer outcome horizons, multi-dimensional responses, and subtype-specific predictors, yet current PD/STN concentration constrains cross-indication generalizability [82]. Yet the principal findings of this review, on external validation rates, TRL distribution, and translational barriers, derive predominantly from STN-targeted PD studies; their applicability to other targets and indications should not be assumed without dedicated replication in those populations.

Accuracy and correlation, while statistically informative, do not reveal how models perform at clinically relevant decision thresholds. For a clinician deciding whether to offer DBS, what matters is not overall accuracy but the balance between missing good candidates (false negatives, patients denied potentially beneficial therapy) and incorrectly recommending surgery to poor candidates (false positives, patients exposed to surgical risk without benefit). Similarly, in targeting, missing the border between STN and substantia nigra carries different clinical consequences than misclassifying intra-STN trajectories. High internal accuracies may decline under external testing because models may learn institution-specific patterns, scanner protocols, surgical techniques, patient selection practices, assessment conventions, rather than generalizable biological signals (***Supplementary Table 1***). Categorical tasks with clear anatomical boundaries preserved high accuracy across centres: contact selection (96-100%, n=51 external cohort) [41,42,48], and STN localization using 7T-trained models (89.4% within 2mm neurosurgical thresholds across institutions, n=80) [83]. In contrast, outcome prediction models demonstrated moderate decline (external AUC 0.76-0.77 versus higher internal AUC 0.87-0.88), and cross-domain transfer showed substantial degradation: STN recognition dropped markedly between databases with different sampling frequencies (25kHz vs 24kHz) [38,84].

Temporal validation is critical where physiological signals are non-stationary: electrode-tissue interfaces evolve through impedance drift and gliosis, neural dynamics shift with disease progression, signal features change with medication cycles and chronic stimulation [85,86]. DBS is lifelong therapy; deployment requires stability across months-to-years. Yet only 8.2 % (n= 6/73) of aDBS studies implement temporal validation. Performance generally declined relative to within-session estimates, with degradation varying by task complexity and temporal gap. Binary classification tasks proved relatively stable: STN/GPi/ViM-LFP wake-sleep classification maintained 85-93% accuracy across 5-7 nights in home environments (mixed cohort: PD, ET, TS) [87]. In contrast, multi-class sleep staging and continuous prediction tasks showed steeper declines. Pallidal LFP sleep classifiers (dystonia, HD, PD) declined from 92% to 76% accuracy across nights [57]. STN-LFP multi-class staging degraded to 58-76% for personalized models and 58-68% for cross-modal DL using chronological time-splits (PD) [88,89]. Gait prediction error increased from 0.159 to 0.174 MAE across clinic visits months apart (STN-LFP, PD) [90], and tremor prediction sensitivity dropped from 100% to 50-100% across sessions ≥1 week apart (surface EMG/accelerometery, PD/ET) [91].

Objective assessment tools similarly lack temporal validation, relying instead on cross-validation or random splits poorly suited to longitudinal monitoring as they cannot capture session-to-session variability or detect drift over time. High reported performance frequently coexists with practices limiting generalizability confidence. Small-sample, high-dimensional settings with incomplete separation between feature selection, optimization, and testing, alongside limited patient-level partitioning, yield optimistic estimates.

Calibration and decision-curve analyses remain underreported, despite being most directly relevant to deployment decisions. Accuracy predominates despite limited interpretability in imbalanced settings, while precision and recall are reported inconsistently, concerning given the asymmetric clinical costs intrinsic to DBS (false negatives deny patients potentially beneficial interventions, while false positives expose them to irreversible surgical risk). Pitfalls extend beyond metric choice to aggregation: averaging across sessions or timepoints without patient-level partitioning obscures inter-individual variability and inflates apparent performance, a risk amplified in the small cohorts characterising this literature [92]. AUROC, while a valid discrimination measure, cannot alone determine whether a model should be used in practice, it requires a smoothed calibration plot and net benefit with decision-curve analysis at the clinically relevant threshold [93]. Extreme reported values (AUC >0.95, accuracy >95%) represent upper-bound estimates under idealized conditions; without calibration and utility data, clinicians cannot assess whether performance aligns with acceptable deployment risk tolerances.

Current systems function as decision-support with clinicians retaining “final authority” over patient selection, targeting, programming, and adaptive control. As closed-loop systems advance toward continuous real-time modulation, stronger safeguards are required, override protocols, anomaly detection, standardized auditing, to prevent mismatches between algorithmic recommendations and clinical judgment. Yet meaningful oversight presupposes proper training and judgment to critically assess outputs, capabilities that AI adoption may paradoxically undermine. Under time pressures, complex targeting or programming protocols risk acceptance at face value, functioning as cognitive shortcuts that disincentivize experience-based learning essential for clinical judgment. The risk of automation bias is evident: overly confident model outputs may prompt clinicians to over-trust recommendations, reducing vigilance and progressively shifting DBS practice toward machine-driven workflows increasingly reliant on models reflecting data quality and biases, precisely when human expertise remains essential to detect errors, incompleteness, or contextual mismatches [94]. Translation appears stalled less because of algorithmic sophistication than because DBS is a biologically heterogeneous, highly dynamic intervention where outcomes depend on many interacting clinical and centre-specific factors, making robust multicentre longitudinal validation difficult and costly. This validation immaturity likely contributes to the scarcity of industry-sponsored studies, given regulatory uncertainty, liability risk, and unclear commercial return without reproducible evidence of generalizability.

The field stands at a translational inflection point: distinguishing technically promising work from deployment-ready evidence requires fundamental shifts in evidence generation, validation, and reporting, not marginal architectural gains. With median TRL 3–4, the bottleneck is not model capacity but reproducible evidence under real-world variability, calibrated risk estimates, transparent handling of class imbalance and small-sample/high-dimensional settings, demonstrated stability across centres and time, governance documentation, and open availability of code or models. This demands: (i) validation designs mirroring clinical deployment, (ii) transparent model lifecycles, and (iii) shared infrastructure with common benchmarks and tooling enabling reproducible, transferable results (***Figure 5 B***).

Reliance on internal cross-validation should be complemented by routine external evaluation on independent centres, with pipelines that fully separate training, feature selection, and testing. For objective assessment tools, programming, and adaptive stimulation, temporal validation using chronologically separated data should be a minimum requirement, as random cross-validation cannot capture non-stationarity due to medication cycles, circadian variation, disease progression, or chronic stimulation effects. Validation should also report performance across centres, devices, and acquisition settings, and explicitly test leakage pathways (for example within-subject overlap or repeated measures) that can inflate estimates in small cohorts.

Beyond technical validation, translation requires operationally useful interpretability. FDA SaMD guidance calls for traceability and human oversight; the EU AI Act classifies DBS as high-risk requiring transparency by design and comprehensive documentation[13,14], requirements operationalized across the full AI lifecycle in the FUTURE-AI international consensus framework, which further distinguishes compliance expectations between research-stage and deployable tools [95].

Yet explainability remains superficial: feature importance, dominant in 59.5 % of studies reporting XAI, provides only global model-level insights, not the decision-specific explanations needed for individual clinical decisions. As models grow more complex, particularly DL architectures, feature importance becomes unreliable. Post-hoc methods (SHAP, LIME, attention), adopted in 5% of studies, offer model-agnostic, individual-level interpretability that high-stakes predictions require. These demand interpretability, demographic robustness, and failure detection largely absent from current prototypes. Post-hoc methods like SHAP plots appear in publications but not clinical interfaces. Deployment requires interface-integrated explanations: planning tools displaying trajectory rationales ("1.2mm STN alignment, 95% HFO confidence; override available"), programming interfaces with biomarker-transparent contact rankings, adaptive dashboards with state-switch explanations. This requires auditable governance: versioning, locked evaluation sets, feature provenance, documented update policies. Interpretability alone is insufficient unless models are evaluated within complete clinical workflows. Future studies should assess these workflows using patient-centred endpoints, programming time, cognitive safety, energy efficiency, quality-of-life, not accuracy alone.

Performance reporting should provide minimum viable metrics: for classification, sensitivity, specificity, positive and negative predictive values, AUC-ROC, and confusion matrices at clinical thresholds; for regression, RMSE, MAE, $R^2$, and residuals, all with confidence intervals. F1 and AUC-PR should be interpreted with caution: both disregard true negatives, are improper at clinically relevant thresholds other than $t=0.5$ or $t=$prevalence and conflate statistical with decision-analytic performance without following decision-theoretic principles[93]. Pre-registered protocols reduce selective reporting. Beyond statistical performance, smoothed calibration plots and decision-curve analysis at the clinically relevant threshold should be minimum requirements for deployment claims; together with AUROC and probability distribution plots, these constitute the core set transforming predictions from abstract scores into actionable risk estimates [93].

Translation requires focus on underrepresented disorders (ET, dystonia), underexplored targets, late-onset complications (ataxia in ET, bradykinesia in dystonia), and multi-target strategies.

Addressing scarcity demands phenotype-stratified models, multi-task learning, and transfer learning with domain adaptation. Federated learning enables multicentre development for rare indications without raw data exchange. Synthetic augmentation may improve robustness during training, but all clinical performance claims require validation on strictly real data with patient-level partitions and external/temporal testing. Studies must report proportion, provenance, and generation assumptions of any synthetic samples [96,97] Critically, "data accessibility" requires shared minimum datasets, consistent labels, explicit metadata (device type, acquisition protocol, stimulation state), and open dictionaries enabling replication. Data availability is further constrained by imaging dependencies that limit participation across centres. Reliance on high-field MRI excludes resource-limited sites and archival standard-scanner data. Algorithms robust to heterogeneous imaging would expand global access and unlock existing multicentre datasets, addressing equity gaps and sample-size constraints simultaneously. This requires training across heterogeneous protocols rather than filtering heterogeneity. Advancing rigor demands cultural shifts: DBS-AI consortia with shared datasets, harmonized protocols, standardized benchmarks. Most papers cite data limitations as primary constraint. The field needs “reference tasks” (targeting segmentation, contact selection, symptom decoding) with common evaluation rules and open baselines. Where sharing faces barriers, federated learning requires shared schemas, labelling conventions, and transparent reporting. Funding mechanisms that prioritize studies demonstrating functional integration and external validation (TRL ≥5) over isolated technical demonstrations could shift investment away from incremental algorithmic refinement, which currently dominates the corpus, toward workflow integration and the longitudinal evidence required for clinical adoption. Encouraging open practices such as reusable code, documented pipelines, and shared benchmarks allows results to compound across research groups rather than remaining isolated proof-of-concept demonstrations. These deployment-focused reforms also prepare the field for the next generation of AI. DBS AI will likely be shaped by foundation models: large-scale self-supervised learning on multimodal data and generative models that can represent uncertainty, missingness, and heterogeneity explicitly. These could enable robust patient-state representations, cross-centre generalization through pretraining, and clinician-facing decision support. However, generative and agentic systems also amplify risks, dataset shift, opaque failures, automation bias, uncontrolled updates, making governance, versioning and auditable evaluation more critical. As tools emerge, "future readiness" should become a scientific standard: open pipelines, benchmark tasks, locked evaluation sets, external/temporal validation as prerequisites for clinical claims rather than afterthoughts. The reforms outlined above are not only necessary for today's models but also enable infrastructure required to safely harness next-generation AI in DBS.

This review reflects inherent challenges in current AI-DBS research. Included studies are highly heterogeneous in design, cohort size, and evaluation strategies, with inconsistently defined metrics limiting direct comparison and precluding meta-analysis. Few studies report calibration, decision-curve analyses, or clinically interpretable thresholds.

Patient demographic characteristics, including age, sex, ethnicity, and socioeconomic factors, were inconsistently reported across studies, precluding assessment of sample representativeness or potential demographic influences on AI model performance and generalizability. Similarly, patient acceptability of AI-driven clinical decision support was not systematically evaluated in the included literature, leaving critical questions regarding trust, usability, and preference for algorithmic versus clinician-led care unanswered.

Scope was restricted to MDs, potentially excluding transferable methodologies from psychiatric or pain applications. TRL assignments relied on evidence within AI-focused publications; downstream clinical trials may not be captured when AI components are not emphasized in indexing. Consequently, TRL estimates may be conservative, with actual translational maturity potentially higher across the broader clinical literature. More broadly, adapting a framework originally designed for engineering systems to the heterogeneous landscape of AI in healthcare involves inherent boundary judgements; while we applied the rubric systematically and with inter-rater adjudication, a domain-specific readiness framework purpose-built for AI-based clinical decision support tools remains an unmet methodological need and a natural extension of this work.

# Methods

This systematic review was conducted and reported in accordance with the PRISMA 2020[98,99] statement and prospectively registered with PROSPERO (CRD42023488688). Following initial data extraction, the analytical framework was expanded beyond the pre-registered protocol to incorporate governance, deployment readiness, and generalisation variables whose relevance and extractability became apparent through iterative engagement with the corpus; this deviation was driven by the exploratory nature of the evidence base rather than by post-hoc outcome considerations.

## Eligibility Criteria

We considered studies applying AI in the context of DBS for MDs, primarily in individuals with PD, ET, or dystonia. Eligible studies used AI to support any stage of the clinical workflow, including preoperative assessment and targeting, postoperative programming, adaptive stimulation, automated clinical evaluation, or to address mechanistic questions directly relevant to DBS action or optimization. AI was defined broadly to include three methodological families: (i) ML, including DL; (ii) symbolic AI, encompassing rule-based approaches; and (iii) fuzzy-logic-based intelligent systems. Throughout this review, 'AI' is used as a field-level umbrella term encompassing all three eligible methodological families. Where claims apply specifically to ML or DL architectures, as verified against the underlying studies, the more specific term is used. Studies were required to report a clearly described data-driven AI model with sufficient detail to identify the learning paradigm, model structure, and validation strategy. Eligible designs included clinical trials, observational and retrospective studies, proof-of-concept investigations, modelling studies using real or realistically simulated data, and preclinical animal studies with direct clinical relevance. We excluded reviews, editorials, conference abstracts without full peer-reviewed publications, and studies in which AI had no direct analytical or decision-support role. Only English-language articles published between 1 January 2000, and 20 April 2025 were considered.

## Search strategy and selection criteria

A structured literature search was conducted in five electronic databases: PubMed, Scopus, Embase, Web of Science, and IEEE Xplore. The search strategy combined controlled vocabulary terms (e.g., MeSH in PubMed and Emtree in Embase) with free-text keywords across three conceptual domains: “Artificial Intelligence”, “Deep Brain Stimulation”, and “Movement Disorders”. Within each domain, synonyms and related terms were combined using the Boolean operator OR, and the three domains were subsequently intersected using AND. The full search strategies for each database are available on Figshare (doi: 10.6084/m9.figshare.31750372) to ensure reproducibility. All retrieved records were imported into a reference management and screening platform (Rayyan Systems Inc., Cambridge, MA, USA) and de-duplicated. Titles and abstracts were screened independently by two reviewers (Z.S. and M.M.U.R.), followed by independent full-text screening of potentially eligible articles. The same dual-reviewer procedure was applied at all subsequent steps requiring independent appraisal (data extraction and TRL assignment): disagreements were resolved through discussion, and where consensus was not reached, by arbitration by a senior reviewer (X.V.).

## Data Extraction and Synthesis

Data extraction followed a piloted, standardised form applied consistently across all 239 eligible studies. Extracted variables captured: (i) general study characteristics, (ii) movement disorder indication, DBS target(s), and clinical context, (iii) workflow stage addressed, (iv) data modalities, (v) AI paradigm and architecture, (vi) validation strategy, and (vii) performance metrics. Beyond these standard variables, extraction was designed to capture translational and deployment-relevant dimensions underreported in prior reviews. Validation independence was assessed using three non-interchangeable categories: external validation required testing on an independent cohort from a different institution than the training data; temporal validation required chronological data splitting or longitudinal follow-up evaluating performance stability across time; and prospective validation required forward-in-time data collection under a pre-specified protocol, distinct from prospective study design alone. A study could satisfy one criterion without satisfying the others, and this distinction is preserved throughout the synthesis. Methodological rigour variables included confidence interval reporting, class imbalance handling, and stratified or subgroup analysis, extracted to assess whether performance claims were adequately qualified for heterogeneous clinical populations. Governance and deployment readiness variables captured whether studies mentioned human-in-the-loop or clinical workflow integration, regulatory or compliance considerations, model lifecycle management practices (locking, versioning, recalibration or retraining protocols), multi-centre data sharing, and open availability of code, model weights, or data. These variables were extracted not as quality scores but as a structured characterisation of how far each study engaged with the conditions required for safe clinical deployment.

Given substantial heterogeneity in patient populations, data modalities, tasks, outcome definitions, and evaluation designs, quantitative pooling was inappropriate and meta-analysis was not pursued. Instead, we conducted qualitative synthesis complemented by descriptive statistics to summarise the distribution of AI methods, data types, validation practices, and performance across the DBS workflow. Studies were grouped by movement disorder, DBS target, workflow stage, and AI model family to identify methodological patterns, recurrent biomarker choices, and differences in translational maturity. This approach prioritised robust, reproducible signals and clearly defined gaps over aggregation of non-comparable effect estimates.

## Risk of bias and reporting quality

Given the methodological heterogeneity of the corpus, encompassing image segmentation, intraoperative signal classification, reinforcement learning, prediction modelling, and mechanistic computational modelling, no single standardised risk-of-bias instrument was applicable across all included designs. PROBAST, the established tool for diagnostic and prognostic prediction-model studies, assumes a defined predictor set, a defined clinical outcome, and patient-level analysis, and therefore does not transfer to a substantial portion of the corpus (e.g., voxel- or trajectory-level evaluation, simulation-based reinforcement learning, mechanistic modelling without a clinical decision endpoint). No alternative single instrument covers the combined methodological landscape of the included studies.

We therefore conducted a structured qualitative appraisal applied uniformly to all 239 studies, in which both reviewers independently assessed each study against a pre-specified set of bias-relevant elements selected to capture the methodological threats most consequential for AI in DBS research: (i) sample size relative to model complexity and presence of class imbalance; (ii) patient-level separation between training and test sets; (iii) presence and independence of external validation; (iv) availability of code, trained models, or data. The resulting study-level appraisals were retained as a structured characterisation of methodological strengths and limitations rather than aggregated into a composite risk-of-bias score, given that no validated weighting scheme exists across the heterogeneous designs of the included corpus.

## Technology readiness level assessment

To quantify translational maturity, we assessed each study using machine-learning-adapted Technology Readiness Levels (TRL 1-9, **Table 5**) [100]. TRL assignment were based solely on information explicitly reported in each study: TRL 1–2 for conceptual/in-silico development; TRL 3 for proof-of-principle in testbed conditions approximating target use; TRL 4 for proof-of-concept on retrospective clinical datasets with internal validation; TRL 5 for systems with evidence of usability beyond offline R&D (e.g., clinician-facing prototypes, shadow mode); TRL 6 for evaluation on prospectively collected data with stated protocols and clinician interaction; and TRL 7–9 for pilot deployment (TRL 7), routine institutional use (TRL 8), or multicentre adoption/regulatory clearance (TRL 9). Both reviewers independently rated each study based on data provenance (retrospective versus prospective, single- versus multicentre), clinical workflow integration, and external validation or deployment evidence; when reporting was insufficient to discriminate adjacent levels, the lower TRL was conservatively assigned. TRL scores reflect translational and deployment maturity, the degree to which a system has been validated in conditions approaching routine clinical use and are independent of scientific quality or mechanistic depth; lower-TRL studies may provide substantial methodological or mechanistic insight while remaining early in the translational pathway. The full scoring rubric, anchoring examples, and case-by-case adjudication rationale are available on Figshare (doi: 10.6084/m9.figshare.31750372)

# Data Availability

All data underlying this review, including the complete extracted dataset for the 239 included studies, the curated datasets used for analysis, definitions, the tables underlying the figures, the code used to generate the figures and statistical analyses, the full TRL scoring rubric, anchoring examples, case-by-case adjudication rationale, and the full search strategies for each database are publicly available on Figshare (doi: 10.6084/m9.figshare.31750372).

# Acknowledgements

No funding was received for this research.

# Author contributions

Z.S. and M.M.U.R. contributed equally and share first authorship. X.V. and L.C. jointly supervised this work and share senior authorship.
Z.S., X.V., and L.C. conceptualized and designed the study. Z.S. and M.M.U.R. performed literature screening and data extraction. Z.S. wrote the first draft of the manuscript. X.V. and L.C. contributed to writing the manuscript. M.M.U.R. generated the figures.

H.A., J.B., S.C., A.F., M.H., J.K.K., A.A.K., P.L., D.E.L., E.M.M., E.M., M.M.R., and L.Z. contributed to interpretation of the results and critically revised the manuscript. All authors reviewed and approved the final manuscript.

# Competing interests

The authors declare no competing financial or non-financial interests.

# Figure legends

### Figure 1: PRISMA flow diagram

### Figure 2: Landscape of AI application across DBS stages

**(A) Distribution of movement disorder** indications across the 239 included studies.

**(B) AI learning paradigm distribution by DBS workflow stage**. Supervised learning dominates across all stages; reinforcement learning is concentrated in adaptive DBS, reflecting its suitability for closed-loop optimization.

**(C) Model architecture group mapped to DBS workflow stage.** Bubble chart mapping model architecture groups to DBS workflow stages (bubble area proportional to number of studies; multi-label counting). Classical approaches, linear models, kernel-based, and tree-based methods, are broadly distributed across workflows, while deep learning architectures (CNNs, feedforward networks) concentrate in targeting and adaptive DBS. Probabilistic and Bayesian models appear most prominently in targeting, consistent with their utility for uncertainty-aware intraoperative decision-making. aDBS, adaptive deep brain stimulation; DBS, deep brain stimulation; GLM, generalized linear model.

### Figure 3: Explainability practices and validation strategies across the DBS workflow

**(A) XAI adoption by workflow stage**.

**(B) XAI method types by workflow stage**; feature importance dominates, post-hoc methods (SHAP, attention) remain marginal.

**(C) Validation strategy distribution by workflow stage**; k-fold cross-validation predominates, temporal validation is negligible.

**(D) Validation types by workflow stage**; internal validation vastly outnumbers external and prospective approaches across all workflow stages.

*aDBS, adaptive deep brain stimulation; CV, cross-validation; LOOCV, leave-one-out cross-validation; LOSO, leave-one-subject-out; SHAP, SHapley Additive exPlanations; XAI, explainable artificial intelligence.*

### Figure 4: Technology readiness level of AI applications across DBS workflow stages

**(A) Technology readiness level mapped to workflow stage.** (N=239) Most stages are dominated by TRL 4 studies. No reviewed system reached routine clinical deployment (TRL ≥7). Systems with prospective clinical evaluation (TRL 6) represented only 5.8% of the full corpus, concentrated in programming (11.1%), aDBS (9.6%), and targeting (6.4%); all remaining systems operate exclusively under research conditions.

**(B) Symptom–signal mapping distribution of adaptive deep brain stimulation research papers across input features, target symptoms, and Technology Readiness Levels**

Each dot represents one research paper. Papers are organized along the x-axis by input bio signal feature, local field potentials (LFP), electromyography (EMG), electrocorticography/electroencephalography/magnetoencephalography (ECoG/EEG/MEG), and biomechanical signals - and along the y-axis by the clinical symptom or functional domain targeted. Dot color reflects the TRL of the study, ranging from TRL 2 (basic research, light blue) to TRL 6 (system demonstration in relevant environment, dark blue). Multiple dots within a cell indicate distinct studies sharing the same feature–symptom combination. Papers utilizing more than one input feature category appear in multiple columns. The figure highlights the dominance of LFP-based approaches for tremor and motor state control, the relative scarcity of higher-TRL validation across most feature–symptom combinations, and underexplored areas warranting future investigation. Biomechanical signals refer to motion-related sensing modalities such as IMU, Kinematics, Accelerometer, Gyroscope etc.

*aDBS, adaptive deep brain stimulation; ECoG, electrocorticography; EEG, electroencephalography; EMG, electromyography; ET, essential tremor; GPi, globus pallidus interna; IMU, inertial measurement unit; LFP, local field potential; PD, Parkinson's disease; STN, subthalamic nucleus; TRL, Technology Readiness Level; Vim, ventral intermediate nucleus.*

## Figure 5: Translational landscape of AI-driven DBS: loop architecture, methodological gaps, and reform pathways

**(A**) **Loop maturity continuum in adaptive deep brain stimulation**. Four progressive stages from fixed-parameter open-loop DBS to therapeutic BCI, mapped to representative control signals. The current clinical frontier — real-time closed-loop aDBS — concentrates 80.8% of the 73 reviewed aDBS studies at TRL 2–4, with only 14 studies (19.2%) reaching TRL 5–6.

**(B**) **From methodological gaps to translational benchmarks.** Four critical gaps identified across 239 studies: validation insufficiency, incomplete reporting, infrastructure void, and disease inequity, mapped to required evidence-based shifts and operational benchmarks for clinical deployment. The overarching transition required is from model architecture optimisation toward reproducible evidence generation under real-world variability, targeting TRL ≥5.

*aDBS, adaptive deep brain stimulation; AUC, area under the curve; BCI, brain–computer interface; CI, confidence interval; EMG, electromyography; IMU, inertial measurement unit; LFP, local field potential; LIME, Local Interpretable Model-agnostic Explanations; PD, Parkinson's disease; SHAP, SHapley Additive exPlanations; STN, subthalamic nucleus; TRL, Technology Readiness Level; XAI, explainable artificial intelligence.*

# Tables

## Table 1. Study characteristics (N= 239)

| Characteristic | Category | n (%) |
|---|---|---|
| ***Disorder category*** | | |
| | PD | 191 (79.91%) |
| | ET | 17 (7.11%) |
| | Dystonia | 10 (4.18%) |
| | Mixed / multiple disorders | 21 (8.78%) |
| ***DBS target*** | | |
| | STN | 147 (61.50%) |
| | GPi | 14 (5.85%) |
| | Vim | 14 (5.85%) |
| | Mixed / multiple targets | 44 (18.41%) |
| | Other[1] | 8 (3.34%) |
| | Not reported / unclear | 12 (5.02%) |
| ***Data modality*** | | |
| | MER | 34 (14.22%) |
| | LFP/ECoG | 29 (12.13%) |
| | MRI/Imaging | 21 (8.78%) |
| | Kinematic/Sensor | 22 (9.20%) |
| | EEG/MEG | 7 (2.92%) |
| | Clinical scores only | 5 (2.09%) |
| | Simulated/Computational | 19 (7.94%) |
| | Multimodal | 100 (41.84%) |
| | Other[2] | 2 (0.83%) |
| ***AI learning paradigm*** | | |
| | Supervised | 199 (83.26%) |
| | Unsupervised | 9 (3.76%) |
| | Reinforcement learning | 16 (6.69%) |
| | Self-supervised | 1 (0.41%) |
| | Mixed paradigms[3] | 11 (4.60%) |
| | Not applicable[4] | 3 (1.25%) |
| ***Validation approach (mutually exclusive primary category)*** | | |
| | Internal validation only | 183 (76.56%) |
| | External validation only | 4 (1.67%) |
| | Prospective validation only | 1 (0.41%) |
| | Multiple validation types[5] | 25 (10.46%) |
| | No formal validation | 26 (10.87%) |
| ***Validation type (non-exclusive; studies may appear in multiple rows)*** | | |
| | Internal validation | 207 (86.61%) |
| | External validation | 14 (5.85%) |
| | Temporal validation | 6 (2.51%) |
| | Prospective validation | 13 (5.43%) |
| ***AI model family*** | | |
| | Traditional Machine Learning | 147 (61.50%) |
| | Deep Learning | 58 (24.26%) |
| | Both (Traditional ML + Deep Learning) | 31 (12.97%) |
| | Not applicable[6] | 3 (1.25%) |

*Values are presented as n (%). Percentages are calculated using N= 239*

*[1] Other DBS targets: CnF/optogenetic (n=1), thalamus (n=2), PPN (n=1), PSA/DRTT (n=1), basal ganglia unspecified (n=2), Gpm (n=1).*

*[2] Other data modality: DBS stimulation parameters (n=1), behavioral task data (n=1).*

*[3] Mixed AI paradigms: Supervised + Unsupervised (n=9), Supervised + Semi-supervised (n=1), Supervised + Self-supervised (n=1).*

*[4] Not applicable AI paradigm: Rule-based (n=2), control-theoretic/adaptive estimation (n=1)*

*[5] Multiple validation types breakdown: Internal + Prospective (n=9), Internal + External (n=7), Internal + Temporal (n=6), Internal + External + Prospective (n=2), External + Prospective (n=1).*

*[6] Not applicable model family: Control-theoretic/fuzzy logic (n=2), control systems engineering (n=1)*

*GPi count (n=14) includes: 12 explicit GPi entries, EP/entopeduncular nucleus (rodent GPi homolog, n=1), and GPi/BG (n=1). These are anatomical equivalents.*

## Table 2. AI model families in deep brain stimulation: A clinical guide

| AI model family | How it works — For clinicians | Key DBS applications | Strengths | Main limitations |
|---|---|---|---|---|
| **Classical Machine Learning** (Linear/GLM, SVM, Random Forest, k-NN) | Trained on manually engineered signal features (e.g., beta power, spike rate). Learns decision boundaries separating tissue types or clinical outcome groups. | MER tissue/structure classification; outcome prediction; contact selection | Interpretable; performs well with small datasets; most externally validated model family in current DBS literature | Requires expert feature engineering; may miss complex nonlinear relationships; performance gains plateau as dataset size increases |

| **Deep Learning** (CNN, Feedforward NN, LSTM) | Automatically learns hierarchical features directly from raw inputs (images, spectrograms, time-series) without manual feature extraction. | Image-based targeting; MER/LFP analysis; adaptive state detection | High performance on complex pattern-recognition tasks; end-to-end learning from raw data | Requires large datasets; limited interpretability ("black box"); higher overfitting risk in small cohorts typical of DBS studies |
|---|---|---|---|---|
| **Probabilistic / Bayesian Models** (HMM, Gaussian Process, Bayesian Networks) | Explicitly models uncertainty and temporal dependencies. Updates probability estimates of neural structure identity along the surgical trajectory. | Intraoperative boundary detection; trajectory planning; uncertainty-aware targeting | Provides calibrated confidence estimates; trajectory-aware; outputs clinically actionable probabilities | Computationally intensive; requires careful prior specification; underused despite clinical relevance of uncertainty quantification |
| **Reinforcement Learning** (Q-learning, Actor-Critic, PPO) | Learns optimal stimulation policies through trial-and-reward interaction with simulated or real physiological feedback. Does not require labelled datasets. | Adaptive stimulation policies; multi-objective programming; closed-loop parameter tuning | Dynamically adapts over time; can optimize long-term outcomes rather than single timepoints | Mostly validated in simulation; safety constraints unresolved for real-time clinical deployment; limited temporal validation in human cohorts |
| **Unsupervised / Generative Models** (Autoencoders, VAE, GAN, Clustering) | Identifies latent structure without outcome labels. Can generate synthetic samples to augment limited datasets. | Biomarker discovery; data augmentation for rare indications (e.g., dystonia); anomaly detection | No labelled data required; reveals hidden patterns; useful for underrepresented conditions | Outputs difficult to clinically validate; synthetic data cannot replace real external validation for clinical performance claims |

*AI = Artificial Intelligence; DBS = Deep Brain Stimulation; GLM= Generalized Linear Model; SVM= Support Vector Machine; k-NN= k-Nearest Neighbours; CNN= Convolutional Neural Network; NN= Neural Network; LSTM= Long Short-Term Memory; HMM= Hidden Markov Model; PPO= Proximal Policy Optimization; MER= Microelectrode Recording; LFP= Local Field Potential; VAE= Variational Autoencoder; GA = Generative Adversarial Network.*

## Table 3. Performance metrics reporting, validation strategies and rates by workflow stage

| Workflow stage | Typical unit of analysis | Predominant validation strategy | External validation rate | Commonly reported performance metrics | Clinical need acuity | Workflow integration mentioned | Key limitations for clinical translation |
|---|---|---|---|---|---|---|---|
| **Targeting** | Trajectory / image volume | Predominantly internal cross-validation | 2/ 62 | Accuracy, sensitivity/specificity, precision/recall/F1 | High: reduces experience-dependent MER analysis and OR time; well-defined task with anatomical ground truth | 69.4% (43/62) | Non-patient-level analysis inflates performance; limited external and temporal validation; uncertainty and robustness rarely reported |
| **Outcome prediction** | Patient | Internal cross-validation (including nested CV) | 3/ 36 | Sensitivity/specificity, correlation, AUC, accuracy | Moderate: genuine patient selection gap but prediction target is heterogeneous and outcome horizons exceed model scope | 50.0% (18/36) | Class imbalance often insufficiently addressed; PR-AUC and calibration rarely reported; limited stratified robustness |

| | | | | | | | across cohorts |
|---|---|---|---|---|---|---|---|
| **Programming** | Patient / visit | Internal cross-validation | 4/ 27 | Accuracy, sensitivity/specificity, AUC, correlation | High: combinatorial explosion ($10^{10}$–$10^{30}$ configurations) is intractable manually; reduces specialist access barrier | 77.8% (21/27) | Small and heterogeneous cohorts; lack of temporal validation; models rarely evaluated in real-world programming workflows |
| **Adaptive DBS** | Signal segment / NA | Internal cross-validation and hold-out | 2/ 73 | Accuracy, sensitivity/specificity | High: addresses fundamental limitation of continuous stimulation; real-time non-stationarity technically demanding | 39.7% (29/73) | Offline evaluation only; absence of external validation; no prospective or closed-loop clinical testing |
| **Objective assessment** | Segment / patient | Predominantly internal cross-validation | 3/ 19 | Accuracy, AUC, correlation | Moderate–high: captures symptom dimensions invisible to standard scales; enables remote monitoring | 42.1% (8/19) | Label noise and inter-rater variability; limited use of clinically meaningful endpoints; robustness across devices and tasks rarely assessed |
| **Mechanistic modeling** | NA | Descriptive, simulation-based | None | Correlation, accuracy, AUC | Low (translational): primarily hypothesis-generating; no direct clinical decision output | 10.5% (2/19) | No direct clinical ground truth; indirect validation only; limited translational relevance without experimental or clinical confirmation |

## Table 4. Translational readiness indicators

| Indicator | Operational definition | n / 239 (%) |
|---|---|---|
| **Regulatory / compliance considerations mentioned** | The paper explicitly references regulatory, compliance, or quality-management frameworks relevant to medical AI or medical devices (e.g., FDA clearance/approval, CE marking, EU MDR, EU AI Act, ISO 13485, IEC 62304, ISO 14971, Good ML Practice, IDE approval), or discusses a regulatory approval/compliance pathway in concrete terms | 12 (5.02%) |
| **Model lifecycle management (locking, versioning, recalibration/retraining)** | The paper reports model locking or freezing for deployment, version control or versioning strategy, monitoring or performance-drift considerations with a plan for recalibration, explicit retraining schedules, or an update policy for adaptive models | 6 (2.51%) |
| **Multi-centre data or collaboration / data sharing** | The dataset includes data from ≥2 clinical centres, or the study explicitly reports multi-centre collaboration with pooled data or a formal data-sharing arrangement | 35 (14.64%) |
| **Open availability (code, model, or data)** | At least one of the following is publicly available and clearly linked or described: source code repository, trained model/weights, or dataset (full or partial) with access instructions | 50 (20.92%) |
| **Prospective evaluation reported** | Prospective or real-time evaluation is explicitly described (distinct from retrospective dataset splitting or prospective data collection with retrospective model evaluation) | 15 (6.27%) |
| **Human-in-the-loop / clinical workflow integration mentioned** | The paper mentions or proposes how the model would be used by clinicians (e.g., decision support, user interface, clinical endpoint, interaction with clinician workflow), beyond reporting algorithmic performance metrics | 122 (51.04%) |

## Table 5. ML-adapted technology readiness levels (TRLs) for DBS-AI: Evidence requirements and translational benchmarks

| TRL | Label (MLTRL-aligned) | Primary evidence markers |
|---|---|---|
| **1** | Concept / first principles | Conceptual algorithm, theoretical framework, or feasibility rationale; no requirement for real clinical data |
| **2** | Goal-oriented research | Low-level experiments on sample data (subset, pilot, or synthetic representative data) to test specific properties or hypotheses |
| **3** | Proof-of-principle in testbeds | Demonstration in testbed conditions (simulated, semi-simulated, or controlled environments) approximating the intended clinical use |
| **4** | Proof-of-concept | Model trained and validated on real retrospective clinical data (often single centre), with internal validation methods explicitly described |
| **5** | Capability demonstrated | Evidence the system is usable beyond offline R&D, such as a clinician-facing decision-support prototype, shadow-mode operation, or demonstrated feasibility of integrating outputs into a DBS workflow (internal use only) |
| **6** | Prospective data in workflow context | Evaluation using prospectively collected intraoperative or postoperative data under a stated protocol, with clinician interaction or explicit workflow constraints |
| **7** | Integrated pilot deployment | AI tool integrated into a clinical or technical stack and used in practice as a pilot deployment, beyond a one-off experimental study |
| **8** | Routine use | Explicit statement of routine operational use within a clinical centre or vendor ecosystem |
| **9** | Multicentre adoption and/or regulatory clearance | Explicit multicentre adoption and/or regulatory clearance as part of a medical device |

**Identification of New Studies via Databases**

Identification

Records identified from:
Databases (n = 5):
**Embase via Ovid** (n = 1122)
**Web of Science** (n = 942)
**Scopus** (n = 2192)
**Pubmed** (n = 118)
**IEEE** (n = 108)
**Total (n = 4482)**

Records identified from other sources (Citation search)
**(n = 48)**

Screening

Records removed before screening:
Duplicate records deleted **(n = 1218)**

Records screened
**(n = 3312)**

Records excluded
**(n = 2984)**

Records sought for retrieval
**(n = 328)**

Records not retrieved
**(n = 0)**

Records assessed for eligibility
**(n =328)**

Records excluded:
Not Relevant to Scope of Review **(n = 89)**

Included

Records included in review
**(n = 239)**

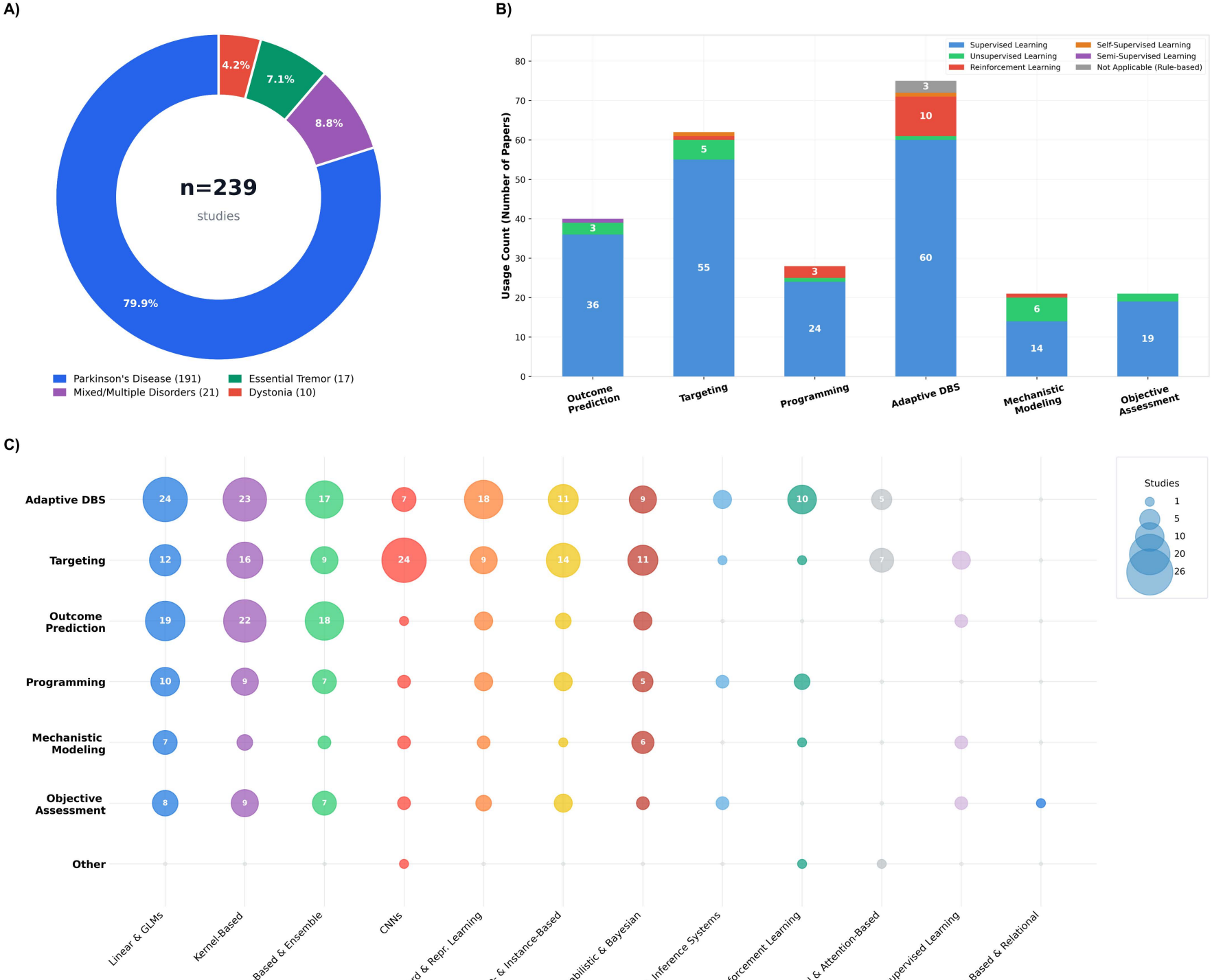

A)
n=239
studies
4.2%
7.1%
8.8%
79.9%
Parkinson's Disease (191)
Essential Tremor (17)
Mixed/Multiple Disorders (21)
Dystonia (10)
B)
Usage Count (Number of Papers)
Supervised Learning
Unsupervised Learning
Reinforcement Learning
Self-Supervised Learning
Semi-Supervised Learning
Not Applicable (Rule-based)
Outcome Prediction
Targeting
Programming
Adaptive DBS
Mechanistic Modeling
Objective Assessment
C)
Adaptive DBS
Targeting
Outcome Prediction
Programming
Mechanistic Modeling
Objective Assessment
Other
Studies
Linear & GLMs
Kernel-Based
Tree-Based & Ensemble
CNNs
Feedforward & Repr. Learning
Distance- & Instance-Based
Probabilistic & Bayesian
Fuzzy Logic & Inference Systems
Reinforcement Learning
Sequential & Attention-Based
Unsupervised Learning
Graph-Based & Relational

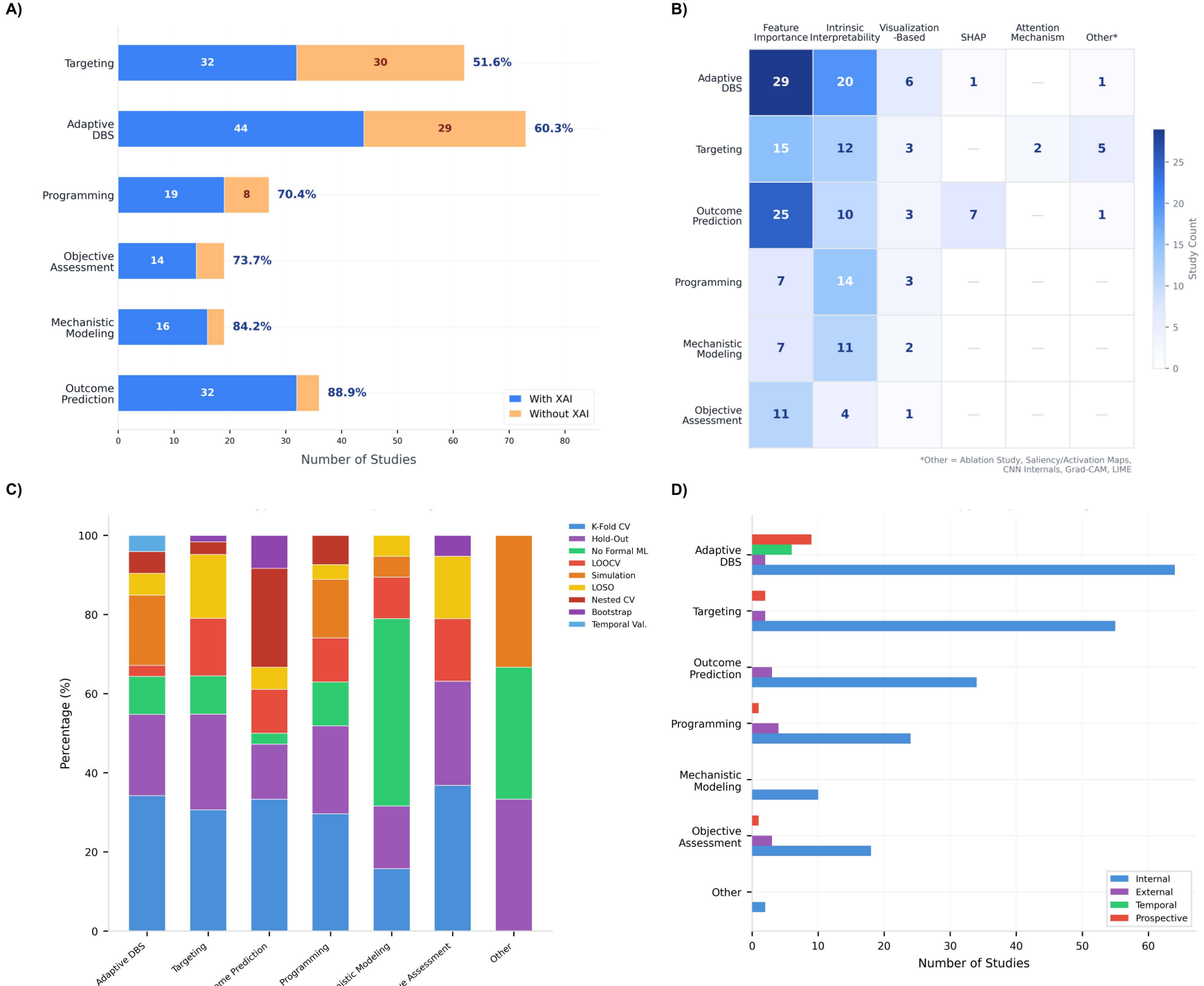

A)
Targeting
Adaptive DBS
Programming
Objective Assessment
Mechanistic Modeling
Outcome Prediction
32
30
51.6%
44
29
60.3%
19
8
70.4%
14
73.7%
16
84.2%
32
88.9%
With XAI
Without XAI
Number of Studies
B)
Feature Importance
Intrinsic Interpretability
Visualization -Based
SHAP
Attention Mechanism
Other*
Study Count
*Other = Ablation Study, Saliency/Activation Maps, CNN Internals, Grad-CAM, LIME
C)
Percentage (%)
K-Fold CV
Hold-Out
No Formal ML
LOOCV
Simulation
LOSO
Nested CV
Bootstrap
Temporal Val.
Adaptive DBS
Targeting
Outcome Prediction
Programming
Mechanistic Modeling
Objective Assessment
Other
D)
Internal
External
Temporal
Prospective
Number of Studies

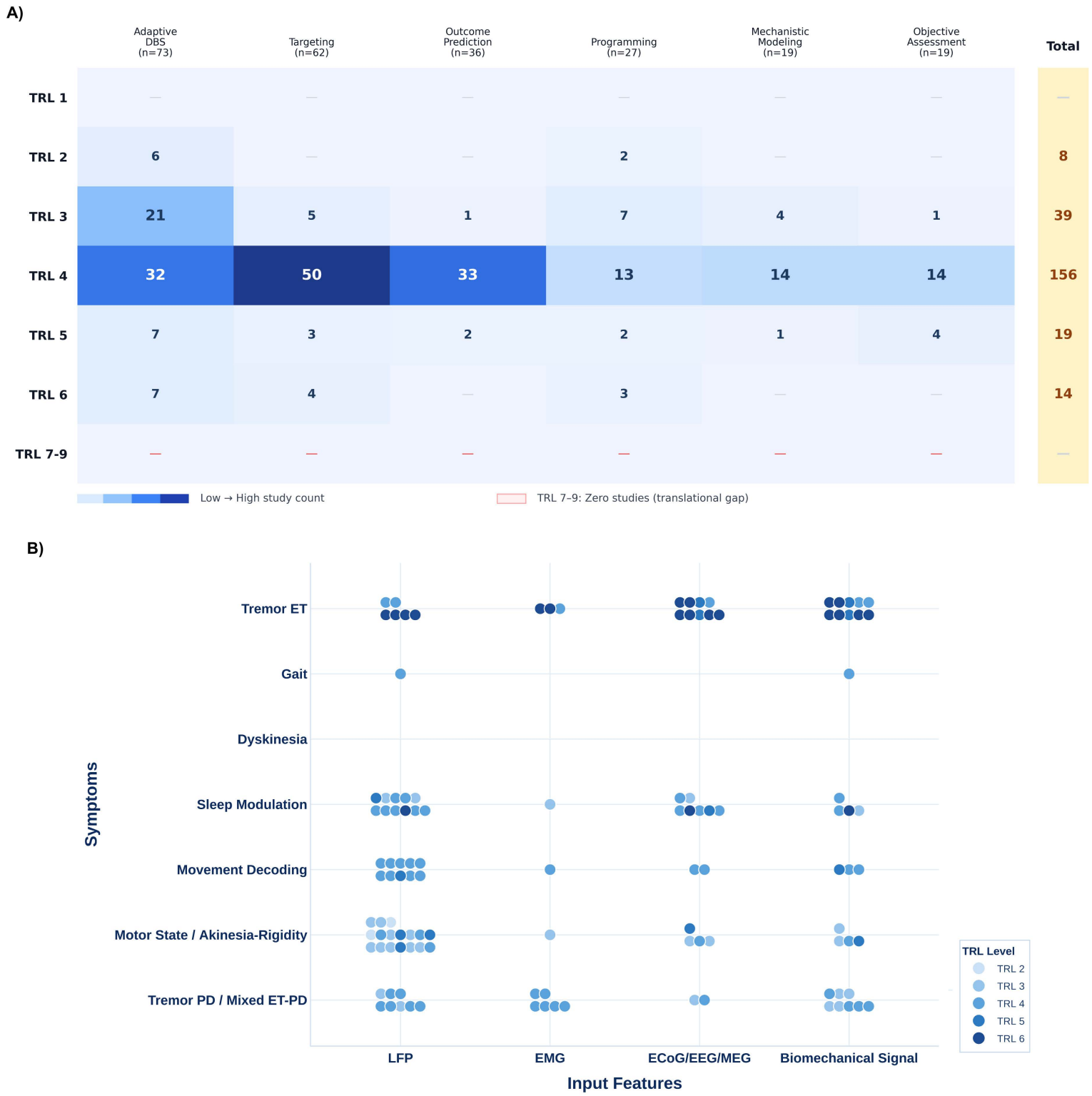

**A)**

| | Adaptive DBS (n=73) | Targeting (n=62) | Outcome Prediction (n=36) | Programming (n=27) | Mechanistic Modeling (n=19) | Objective Assessment (n=19) | **Total** |
|---|---|---|---|---|---|---|---|
| **TRL 1** | — | — | — | — | — | — | — |
| **TRL 2** | 6 | — | — | 2 | — | — | 8 |
| **TRL 3** | 21 | 5 | 1 | 7 | 4 | 1 | 39 |
| **TRL 4** | 32 | 50 | 33 | 13 | 14 | 14 | 156 |
| **TRL 5** | 7 | 3 | 2 | 2 | 1 | 4 | 19 |
| **TRL 6** | 7 | 4 | — | 3 | — | — | 14 |
| **TRL 7-9** | — | — | — | — | — | — | — |

Low → High study count

TRL 7–9: Zero studies (translational gap)

**B)**

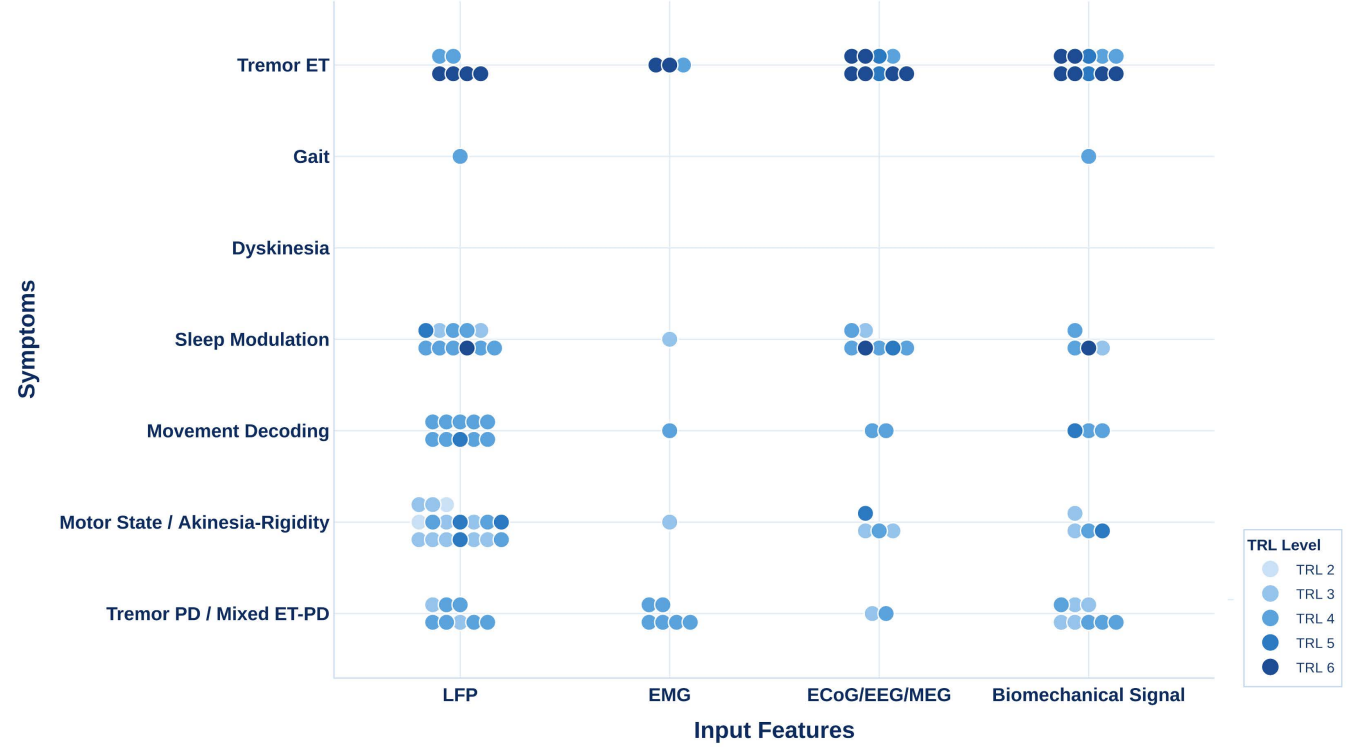

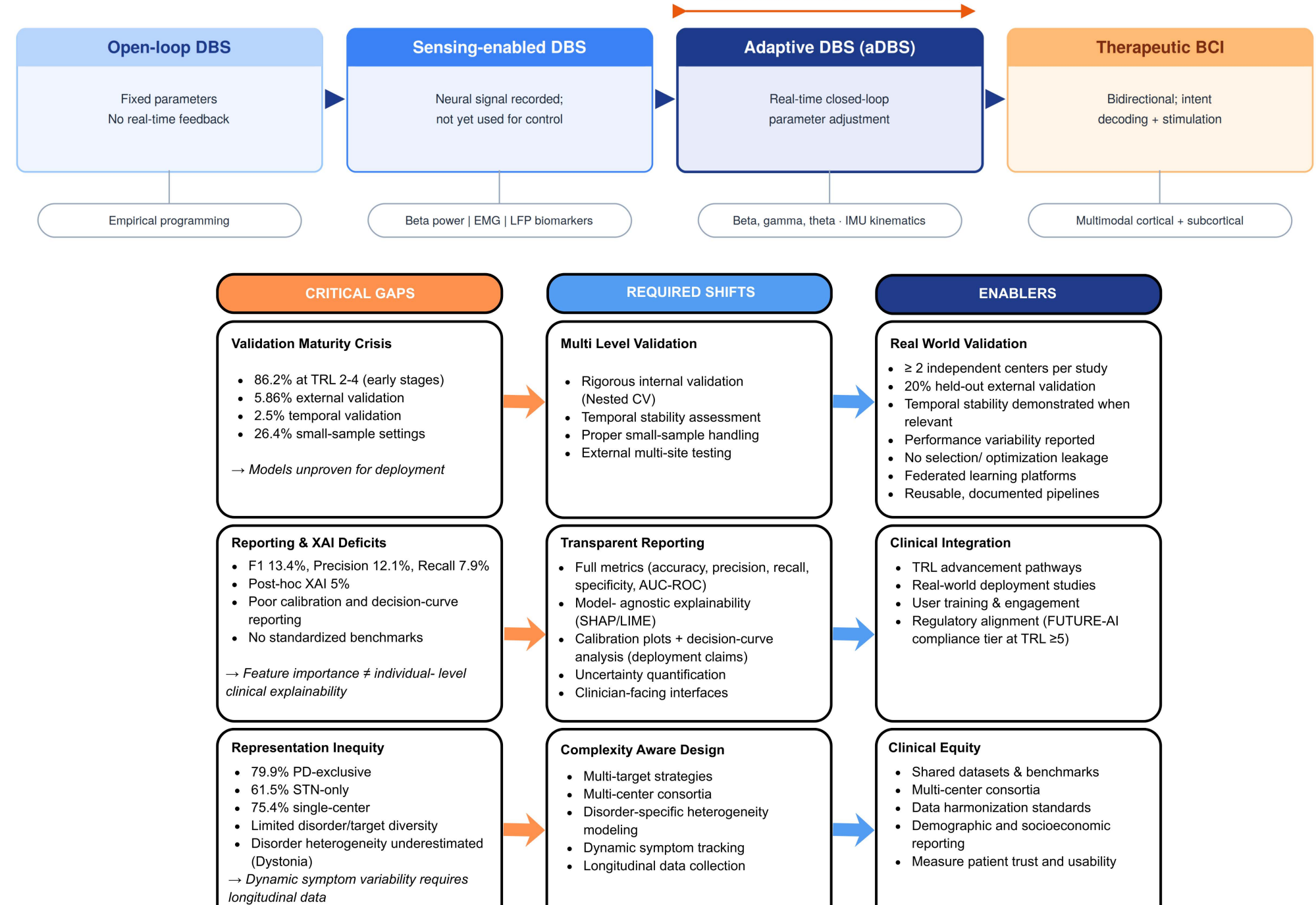

A)
Current Clinical Frontier (7 Studies at TRL 6)
Open-loop DBS
Fixed parameters
No real-time feedback
Empirical programming
Sensing-enabled DBS
Neural signal recorded;
not yet used for control
Beta power | EMG | LFP biomarkers
Adaptive DBS (aDBS)
Real-time closed-loop
parameter adjustment
Beta, gamma, theta · IMU kinematics
Therapeutic BCI
Bidirectional; intent
decoding + stimulation
Multimodal cortical + subcortical
B)
CRITICAL GAPS
REQUIRED SHIFTS
ENABLERS
Validation Maturity Crisis
• 86.2% at TRL 2-4 (early stages)
• 5.86% external validation
• 2.5% temporal validation
• 26.4% small-sample settings
→ Models unproven for deployment
Multi Level Validation
• Rigorous internal validation (Nested CV)
• Temporal stability assessment
• Proper small-sample handling
• External multi-site testing
Real World Validation
• ≥ 2 independent centers per study
• 20% held-out external validation
• Temporal stability demonstrated when relevant
• Performance variability reported
• No selection/ optimization leakage
• Federated learning platforms
• Reusable, documented pipelines
Reporting & XAI Deficits
• F1 13.4%, Precision 12.1%, Recall 7.9%
• Post-hoc XAI 5%
• Poor calibration and decision-curve reporting
• No standardized benchmarks
→ Feature importance ≠ individual- level clinical explainability
Transparent Reporting
• Full metrics (accuracy, precision, recall, specificity, AUC-ROC)
• Model- agnostic explainability (SHAP/LIME)
• Calibration plots + decision-curve analysis (deployment claims)
• Uncertainty quantification
• Clinician-facing interfaces
Clinical Integration
• TRL advancement pathways
• Real-world deployment studies
• User training & engagement
• Regulatory alignment (FUTURE-AI compliance tier at TRL ≥5)
Representation Inequity
• 79.9% PD-exclusive
• 61.5% STN-only
• 75.4% single-center
• Limited disorder/target diversity
• Disorder heterogeneity underestimated (Dystonia)
→ Dynamic symptom variability requires longitudinal data
Complexity Aware Design
• Multi-target strategies
• Multi-center consortia
• Disorder-specific heterogeneity modeling
• Dynamic symptom tracking
• Longitudinal data collection
Clinical Equity
• Shared datasets & benchmarks
• Multi-center consortia
• Data harmonization standards
• Demographic and socioeconomic reporting
• Measure patient trust and usability
Core Paradigm Shift: Model architecture optimization → Reproducible evidence generation under real-world conditions